\documentclass[aoas,MSNbibl,nameyear,dvips]{arximspdf}

\doi{10.1214/13-AOAS689} 
\volume{7}
\issue{4}
\pubyear{2013}
\firstpage{1837}
\lastpage{1837}

\begin{document}
\begin{frontmatter}

\title{Editorial}
\runtitle{Editorial}

\begin{aug}
\author{\fnms{Susan M.} \snm{Paddock}\ead[label=e1]{paddock@rand.org}}
\runauthor{S. M. Paddock}
\affiliation{RAND Corporation}
\address{RAND Corporation\\
1776 Main Street\\
Santa Monica, California 90401\\
USA\\
\printead{e1}} 
\end{aug}

\received{\smonth{9} \syear{2013}}



\end{frontmatter}

Clauset and Woodard (\citeyear{ClaWoo}) ask, ``What is the likelihood of
another September 11th-sized or larger terrorist event, worldwide, over
the next decade?'' This question has implications for numerous policy
arenas---national security, international relations, public safety,
disaster preparedness, and so on---but it also has resonance for any
individual who remembers 9/11. Thus, the Area Editors chose this paper
to engage the audience at the 2013 Joint Statistical Meetings (JSM) in
Montreal. The discussions in this issue of \textit{The Annals of Applied
Statistics} include those presented formally at JSM 2013 as well as
others from individuals who were unable to attend JSM 2013 or who
contributed discussions after the meeting.

Clauset and Woodard (\citeyear{ClaWoo}) did indeed stimulate an active discussion.
Several points were raised in the discussion about assumptions.
Substantive assumptions that were questioned include the framing of the
problem---for example, how useful is it to characterize the impact of
an attack by the number of casualties---and the quality and
informativeness of terrorism data in general. Assumptions about the
statistical modeling strategy were also questioned, particularly with
respect to how best to model rare and extreme events and whether
terrorist events are independent and identically distributed.
Considering the controversial topic and the sensitivity of results to
different models for rare event data, careful consideration of the
assumptions is appropriate. Clauset and Woodard (\citeyear{ClaWoo}) and the
accompanying discussions should provide ample material for an
interested reader to evaluate the importance of these assumptions to
examining whether a 9/11-sized terrorist attack is an outlier.

\def\bibname{Reference}

%



\printaddresses

\end{document}